\catcode`@=11

\newskip\plaincentering \plaincentering=0pt plus 1000pt minus 1000pt
\def\@plainlign{\tabskip=0pt\everycr={}}
\def\eqalignno#1{\displ@y \tabskip\plaincentering
  \halign to\displaywidth{\hfil$\@lign\displaystyle{##}$\tabskip\z@skip
    &$\@lign\displaystyle{{}##}$\hfil\tabskip\plaincentering
    &\llap{$\@lign##$}\tabskip\z@skip\crcr
    #1\crcr}}
\def\leqalignno#1{\displ@y \tabskip\plaincentering
  \halign to\displaywidth{\hfil$\@lign\displaystyle{##}$\tabskip\z@skip
    &$\@lign\displaystyle{{}##}$\hfil\tabskip\plaincentering
    &\kern-\displaywidth\rlap{$\@lign##$}\tabskip\displaywidth\crcr
    #1\crcr}}
\def\plainLet@{\relax\iffalse{\fi\let\\=\cr\iffalse}\fi}
\def\plainvspace@{\def\vspace##1{\noalign{\vskip##1}}}

\def\intic@{\mathchoice{\hskip5\p@}{\hskip4\p@}{\hskip4\p@}{\hskip4\p@}}
\def\negintic@
 {\mathchoice{\hskip-5\p@}{\hskip-4\p@}{\hskip-4\p@}{\hskip-4\p@}}
\def\intkern@{\mathchoice{\!\!\!}{\!\!}{\!\!}{\!\!}}
\def\intdots@{\mathchoice{\cdots}{{\cdotp}\mkern1.5mu
    {\cdotp}\mkern1.5mu{\cdotp}}{{\cdotp}\mkern1mu{\cdotp}\mkern1mu
      {\cdotp}}{{\cdotp}\mkern1mu{\cdotp}\mkern1mu{\cdotp}}}
\newcount\intno@
\def\iint{\intno@=\tw@\futurelet\next\ints@}
\def\iiint{\intno@=\thr@@\futurelet\next\ints@}
\def\iiiint{\intno@=4 \futurelet\next\ints@}
\def\idotsint{\intno@=\z@\futurelet\next\ints@}
\def\ints@{\findlimits@\ints@@}
\newif\iflimtoken@
\newif\iflimits@
\def\findlimits@{\limtoken@false\limits@false\ifx\next\limits
 \limtoken@true\limits@true\else\ifx\next\nolimits\limtoken@true\limits@false
    \fi\fi}
\def\multintlimits@{\intop\ifnum\intno@=\z@\intdots@
  \else\intkern@\fi
    \ifnum\intno@>\tw@\intop\intkern@\fi
     \ifnum\intno@>\thr@@\intop\intkern@\fi\intop}
\def\multint@{\int\ifnum\intno@=\z@\intdots@\else\intkern@\fi
   \ifnum\intno@>\tw@\int\intkern@\fi
    \ifnum\intno@>\thr@@\int\intkern@\fi\int}
\def\ints@@{\iflimtoken@\def\ints@@@{\iflimits@
   \negintic@\mathop{\intic@\multintlimits@}\limits\else
    \multint@\nolimits\fi\eat@}\else
     \def\ints@@@{\multint@\nolimits}\fi\ints@@@}
\def\Sb{_\bgroup\vspace@
        \baselineskip=\fontdimen10 \scriptfont\tw@
        \advance\baselineskip by \fontdimen12 \scriptfont\tw@
        \lineskip=\thr@@\fontdimen8 \scriptfont\thr@@
        \lineskiplimit=\thr@@\fontdimen8 \scriptfont\thr@@
        \Let@\vbox\bgroup\halign\bgroup \hfil$\scriptstyle
            {##}$\hfil\cr}
\def\endSb{\crcr\egroup\egroup\egroup}
\def\Sp{^\bgroup\vspace@
        \baselineskip=\fontdimen10 \scriptfont\tw@
        \advance\baselineskip by \fontdimen12 \scriptfont\tw@
        \lineskip=\thr@@\fontdimen8 \scriptfont\thr@@
        \lineskiplimit=\thr@@\fontdimen8 \scriptfont\thr@@
        \Let@\vbox\bgroup\halign\bgroup \hfil$\scriptstyle
            {##}$\hfil\cr}
\def\endSp{\crcr\egroup\egroup\egroup}
\def\Let@{\relax\iffalse{\fi\let\\=\cr\iffalse}\fi}
\def\vspace@{\def\vspace##1{\noalign{\vskip##1 }}}
\def\aligned{\,\vcenter\bgroup\plainvspace@\plainLet@\openup\jot\m@th\ialign
  \bgroup \strut\hfil$\displaystyle{##}$&$\displaystyle{{}##}$\hfil\crcr}
\def\endaligned{\crcr\egroup\egroup}
\def\matrix{\,\vcenter\bgroup\plainLet@\plainvspace@
    \normalbaselines
  \m@th\ialign\bgroup\hfil$##$\hfil&&\quad\hfil$##$\hfil\crcr
    \mathstrut\crcr\noalign{\kern-\baselineskip}}
\def\endmatrix{\crcr\mathstrut\crcr\noalign{\kern-\baselineskip}\egroup
                \egroup\,}
\newtoks\hashtoks@
\hashtoks@={#}
\def\format{\crcr\egroup\iffalse{\fi\ifnum`}=0 \fi\format@}
\def\format@#1\\{\def\preamble@{#1}%
  \def\c{\hfil$\the\hashtoks@$\hfil}%
  \def\r{\hfil$\the\hashtoks@$}%
  \def\l{$\the\hashtoks@$\hfil}%
  \setbox\z@=\hbox{\xdef\Preamble@{\preamble@}}\ifnum`{=0 \fi\iffalse}\fi
   \ialign\bgroup\span\Preamble@\crcr}

\def\cases{\left\{\,\vcenter\bgroup\plainvspace@
     \normalbaselines\openup\jot\m@th
      \plainLet@\ialign\bgroup$\displaystyle{##}$\hfil&\quad$\displaystyle{{}##}$\hfil\crcr
      \mathstrut\crcr\noalign{\kern-\baselineskip}}

\newif\iftagsleft@
\tagsleft@true
\def\TagsOnRight{\global\tagsleft@false}
\def\tag#1$${\iftagsleft@\leqno\else\eqno\fi
 \hbox{\def\pagebreak{\global\postdisplaypenalty-\@M}%
 \def\nopagebreak{\global\postdisplaypenalty\@M}\rm(#1\unskip)}%
  $$\postdisplaypenalty\z@\ignorespaces}
\interdisplaylinepenalty=\@M
\def\plainallowdisplaybreak@{\def\allowdisplaybreak{\noalign{\allowbreak}}}
\def\plaindisplaybreak@{\def\displaybreak{\noalign{\break}}}
\def\align#1\endalign{\def\tag{&}\plainvspace@\plainallowdisplaybreak@\plaindisplaybreak@
  \iftagsleft@\plainlalign@#1\endalign\else
   \plainralign@#1\endalign\fi}
\def\plainralign@#1\endalign{\displ@y\plainLet@\tabskip\plaincentering\halign to\displaywidth
     {\hfil$\displaystyle{##}$\tabskip=\z@&$\displaystyle{{}##}$\hfil
       \tabskip=\plaincentering&\llap{\hbox{\rm(##\unskip)}}\tabskip\z@\crcr
             #1\crcr}}
\def\plainlalign@
 #1\endalign{\displ@y\plainLet@\tabskip\plaincentering\halign to \displaywidth
   {\hfil$\displaystyle{##}$\tabskip=\z@&$\displaystyle{{}##}$\hfil
   \tabskip=\plaincentering&\kern-\displaywidth
        \rlap{\hbox{\rm(##\unskip)}}\tabskip=\displaywidth\crcr
               #1\crcr}}

\def\re@#1{\par\hangindent\parindent\indent\llap{#1\enspace}\ignorespaces}
\def\qfootnote#1{\edef\@sf{\spacefactor\the\spacefactor}{}#1\@sf
      \insert\footins{\let\egroup=}\footnotesize 
      \interlinepenalty100 \let\par=\endgraf
        \leftskip=0pt \rightskip=0pt
        \splittopskip=10pt plus 1pt minus 1pt \floatingpenalty=20000
   \smallskip\re@{#1}\bgroup\strut\aftergroup{\strut\egroup}\let\next}
\catcode`\@=\active

\documentstyle[11pt]{article}
\begin{document}

\newcommand{\be}{\begin{equation}}
\newcommand{\ee}{\end{equation}}

\newcommand{\ba}{\begin{equation} \aligned}
\newcommand{\ea}{\endaligned \end{equation}}

\title{
\bf Spinor Field Realizations of $W_{2,6}$ String and $W_{6}$
String }
\author{
{Shu-Cheng Zhao,  Li-Jie Zhang, Yu-Xiao Liu}\\
{}\\
{Institute of Theoretical Physics,} \\
{Lanzhou University, Lanzhou, P. R. China, 730000}\\
{}\\
{E-mail: zhaosc@lzu.edu.cn, zlj173071@163.com, liuyx01@st.lzu.edu.cn} }

\maketitle \vskip 21pt
\begin{center}
\begin{minipage}{120mm}
\normalsize \baselineskip 15pt
 {\begin{center}\bf
Abstract\end{center}}
 \indent In this paper the spinor field BRST charges of the W2,6 string and W6 string are constructed, where the
BRST charges are graded.\\


\end{minipage}
\end{center}

\vskip 21pt

\newpage \normalsize \baselineskip 15pt

\noindent{\bf  I Introduction } \\
\indent As is well known, much work on the $W$ algebra and $W$
string has received a considerable attention since 1990's[1-6].
The $BRST$ method is the simplest way to build a $W$ string by
far. In the work[7], we mentioned that the explicit forms of the
$BRST$ charge for the $W_{4}$ and $W_{5}$ algebra[8,9] are so
complex and it is difficult to be generalized to a general $W_{N}$
string. And we point out that the reason why are in two parts:
because the method is not graded and it just only belongs to the
usual scalar field realizations. At the same time, we found the
methods to construct the spinor field realizations of $W_{2,s}$
strings and $W_{N}$ strings. We assume the $BRST$ charges of the
$W_{2,s}$ strings or $W_{N}$ strings are graded, this Ansatz makes
their $BRST$ charges become very easy. And the exact constructions
of $W_{2,5}$ string and $W_{5}$ string have been obtained[7]. \\
\indent Naturally the computational complexity rises rapidly when
we construct the spinor field realization of $W_{2,s}$ and $W_{N}$
strings with higher '$s$' or '$N$'. It is necessary to give out a
general program since the calculation by hand becomes more
difficult. In fact, we have written the program which can give out
the precise solutions in order to check our Ansatz and compute any
solutions of the spinor realizations of $W_{2,s}$ and $W_{N}$
strings. The spinor field $BRST$ charges of $W_{2,6}$ string and
$W_{6}$ string have been constructed firstly by using it.
Especially we improved the results of $W_{2,4}$ string in the
work[10] and obtained a more general result. Of course, we also
checked the results of $W_{2,3}$ and $W_{2,5}$ string[7,11], and
we only spend several minutes on the calculation of $W_{2,3}$
string in particular. So the Ansatz has been showed again and all
the spinor realizations of $W_{2,s}$ and $W_{N}$ strings can be
obtained theoretically by using our program.\\
    \indent This paper is organized as follows. First we give a review
of the grading $BRST$ method to construct spinor field
realizations of the $W_{2,s}$ strings and $W_{N}$ strings. Then we
introduce the thought of our program which can construct the
$BRST$ charge of $W_{2,s}$ strings. Subsequently, we mainly give
out the spinor field $BRST$ charges of $W_{2,6}$ string and
$W_{6}$ string. And the new result of $W_{2,4}$ string is also
included. Finally,
a brief conclusion is given.\\

\noindent{\bf  II Review of BRST Method about Spinor $W_{2,S}$
Strings and $W_{N}$ Strings}\\
\indent Following Ref. [11], the $BRST$ charge for the spinor
field realization of $W_{2,s}$ strings takes the form: \be
    Q_{B}=Q_{0}+Q_{1},
\ee
\be
    Q_{0}=\oint dz\; c(T^{eff}+T_{\psi}+KT_{bc}+yT_{\beta\gamma}),
\ee
\be
    Q_{1}=\oint dz\; \gamma F(\psi,\beta,\gamma),
\ee
\be
    T_{\psi}=-\frac{1}{2}\partial\psi\psi,
\ee
\be
    T_{\beta\gamma}=s\beta\partial\gamma+(s-1)\partial\beta\gamma,
\ee
\be
    T_{bc}=2b\partial c+\partial bc,
\ee
\be
    T^{eff}=-\frac{1}{2}\eta_{\mu\nu}\partial Y^{\mu} Y^{\nu},
\ee
\noindent where $K,y$ are pending constants and $Y^{\mu}$ is
a multi-spinor. It is easy to verify that the condition
$Q_{0}^{2}=0$ is satisfied for any '$s$', so the nilpotency
condition $Q_{B}^{2}=0$ can be translated into
$Q_{1}^{2}=\{Q_{0},Q_{1}\}=0$. Using these conditions, the precise
form of $F(\psi,\beta,\gamma)$ and exact $y$
can be determined.\\
    \indent We can assume the $BRST$ charge of a general $W_{N}$ string
     is also graded [10], and it can be given as follows:
\be
    Q_{B}=Q_{0}+Q_{1}+\cdots +Q_{N-2}=Q_{0}+\sum_{i=1}^{N-2}Q_{i}.
\ee
    \indent Then the nilpotency condition $Q_{B}^{2}=0$ translates
into
\be
    Q_{0}^{2}=Q_{i}^{2}=\left\{Q_{0},Q_{i}\right\}=0,
\ee
\be
    \left\{Q_{i},Q_{j}\right\}_{i<j}=0.
\ee
    \indent Observing these conditions, we can consider the
equations $Q_{0}^{2}=Q_{i}^{2}=\left\{Q_{0},Q_{i}\right\}=0$
correspond to the case of $W_{2,(i+2)}$ strings for each '$i$', so
$Q_{i}$ should take the case of $W_{2,(i+2)}$ strings directly.
And the correct solution can be selected out by using the
equations $\left\{Q_{i},Q_{j}\right\}_{i<j}=0$.\\
    \indent The $BRST$ charge of a spinor $W_{N}$ string
can be written as follows:
\be
    Q_{B}=Q_{0}+Q_{1}+\cdots +Q_{N-2},
\ee
\be
    Q_{0}=\oint dz\; c(T^{eff}+T_{\psi}+KT_{bc}+\sum_{i=1}^{N-2}y_{i}T_{\beta_{i}\gamma_{i}}),
\ee
\be
    Q_{i}=\oint dz\; \gamma_{i}F_{i}(\psi,\beta_{i},\gamma_{i}),
\ee
\be
    T_{\psi}=-\frac{1}{2}\partial\psi\psi,
\ee
\be
    T_{bc}=2b\partial c+\partial bc,
\ee
\be
    T_{\beta_{i}\gamma_{i}}=(i+2)\beta_{i}\partial\gamma_{i}+(i+1)\partial\beta_{i}\gamma_{i},
\ee
\be
    T^{eff}=-\frac{1}{2}\eta_{\mu\nu}\partial Y^{\mu} Y^{\nu},
\ee \noindent where the ghost fields $b,c,\beta_{i},\gamma_{i}$
are all bosonic and communicating whilst the spinor field $\psi$
has spin $\frac{1}{2}$ and is anti-communicating. They satisfy
$OPEs$: \be
    b(z)c(\omega)\sim\frac{1}{z-\omega}\; ,\;
    \beta_{i}(z)\gamma_{i}(\omega)\sim\frac{1}{z-\omega}\; , \;
    \psi(z)\psi(\omega)\sim  -\frac{1}{z-\omega},
\ee
\noindent in the other case the $OPEs$ vanish.\\

\noindent {\bf III Thought of the Corresponding Program}\\
\indent According to above theory, we should obtain the spinor
field realization of $W_{2,(i+2)}$ strings firstly. So producing a
general program for $W_{2,s}$ strings becomes very important.
Constructing the form of $F(\psi,\beta,\gamma)$ and getting the
value of '$y$' are our main work. The thought of our program for
determining $F(\psi,\beta,\gamma)$ and '$y$' can be described as
follows. Firstly, write out all the possible terms of $F$ with
$\psi,\beta,\gamma$ considering the spin of each term '$s$' and
ghost number zero. Then leave out all the total differential terms
in $\gamma F(\psi,\beta,\gamma)$ since their contribution to
$Q_{1}$ is zero. By using the nilpotency conditions
$Q_{1}^{2}=\{Q_{0},Q_{1}\}=0$, we can obtain corresponding
equations subsequently. Finally, the value of '$y$' and all the
coefficients in $F$ would be obtained by solving these
equations. Thus the form of $BRST$ charge is constructed. \\
\indent Here it is not necessary to list the program, if someone
need it, we would like to offer it at any time.\\

\noindent{\bf IV Exact Spinor Field Constructions of the $W_{2,6}$ String}\\
\indent Apply the grading BRST method and our program, we
obtained the results of the $W_{2,6}$ string as follows.\\
\indent The general form of F is:
$$
\align
 F6(\psi,\beta,\gamma)
    &=f6[1]\beta^6\gamma^6+f6[2]\beta^5\partial\gamma\gamma^4+f6[3]\beta^4(\partial\gamma)^2\gamma^2+f6[4]\beta^3(\partial\gamma)^3+f6[5] \beta^4\gamma^4\partial\psi\psi\\\allowdisplaybreak
    &+f6[6]\beta^3\partial\gamma\gamma^2\partial\psi\psi+f6[7]\beta^2(\partial\gamma)^2\partial\psi\psi+f6[8]\beta^3\gamma^3\partial^2\beta +f6[9]\beta^2\gamma^2\partial^2\beta\partial\gamma\\\allowdisplaybreak
    &+f6[10]\beta\gamma^2\partial^2\beta\partial\psi\psi+f6[11](\partial^2\beta)^2\gamma^2+f6[12]\beta^4\partial^2\gamma\gamma^3+f6[13]\beta^3\partial^2\gamma\partial\gamma\gamma\\\allowdisplaybreak
    &+f6[14]\partial\beta\beta\partial^2\gamma\partial\gamma+f6[15]\beta^2\partial^2\gamma\gamma\partial\psi\psi +f6[16]\partial\beta\partial^2\gamma\partial\psi\psi+f6[17]\partial^2\beta\beta\partial^2\gamma\gamma\\\allowdisplaybreak
    &+f6[18]\beta^2(\partial^2\gamma)^2+f6[19]\beta^3\gamma^3\partial^2\psi\psi+f6[20]\beta^2\partial\gamma\gamma\partial^2\psi\psi+f6[21]\beta^2\gamma^2\partial^2\psi\partial\psi\\\allowdisplaybreak
    &+f6[22]\beta\partial\gamma\partial^2\psi\partial\psi+f6[23]\partial^2\beta\gamma\partial^2\psi\psi+f6[24]\beta\partial^2\gamma\partial^2\psi\psi+f6[25]\beta^2\gamma^3\partial^3\beta\\\allowdisplaybreak
    &+f6[26]\partial ^3\beta\beta\partial\gamma\gamma+f6[27]\partial^3\beta\gamma\partial\psi\psi+f6[28]\partial^3\beta\partial^2\gamma +f6[29]\beta^3\partial^3\gamma\gamma^2\\\allowdisplaybreak
    &+f6[30]\beta^2\partial^3\gamma\partial\gamma+f6[31]\beta\partial^3\gamma\partial\psi\psi+f6[32]\partial^2\beta\partial^3\gamma+f6[33]\beta^2\gamma^2\partial^3\psi\psi\\\allowdisplaybreak
    &+f6[34]\beta\partial\gamma\partial^3\psi\psi+f6[35]\beta\gamma\partial^3\psi\partial\psi+f6[36]\partial^3\psi\partial^2\psi+f6[37]\partial^4\beta\beta\gamma^2\\\allowdisplaybreak
    &+f6[38]\beta^2\partial^4\gamma\gamma+ f6[39]\partial\beta\partial^4\gamma+f6[40]\beta\gamma\partial^4\psi\psi +f6[41]\partial^4\psi\partial\psi\\\allowdisplaybreak
    &+f6[42]\partial^5\beta\gamma+f6[43]\beta\partial^5\gamma+f6[44]\partial^5\psi\psi.
\endalign $$
\noindent Three constructions have been worked out as follows:\\
\noindent(1) $y=0$ and \\
\indent  $ f6[5]=f6[6]=f6[7]=f6[10]=f6[15]=f6[19]=f6[20]=f6[21]=f6[33]=f6[36]\\
\indent   =f6[41]=f6[44]=0,\indent f6[16]=f6[22]=f6[24]=f6[34]=C,\indent f6[23]=-\frac{1}{2} C,\\
\indent  f6[27]=-\frac{1}{6} C,\indent f6[31]=f6[35]=\frac{2}{3} C,\indent f6[40]=\frac{1}{3} C,\\
        $
\noindent where $C$ and other coefficients are arbitrary constants but do not vanish at the same
time.\\[2mm]
\noindent (2) $y=1$\\ The forms of $f6[i](i=1,2,...,44)$ are complex comparatively. We do not list this result here.\\[2mm]
\noindent (3) $y$ is an arbitrary constant and\\
\indent $  f6[5]\;=f6[6]=f6[7]=f6[10]=f6[15]=f6[19]=f6[20]=f6[21]=f6[33]=f6[36]\\
 \indent   =f6[41]=f6[44]=0,\indent f6[1]\;\;=16(-302879179L_{2}+102793785L_{3})/281288882105,\\
    \indent f6[2]\;\;=\frac{231}{5}f6[1], \indent f6[3]\;\;=2(-379632019L_{2}+102044487L_{3})/66420043,\\
     \indent f6[4]\;\;=242(-58153384L_{2}+10140597L_{3})/332100215,\\
   \indent f6[8]\;\;=2(2065520459L_{2}-5774235L_{3})/3653102365,\\
   \indent f6[9]\;\;=9(168697384L_{2}-384949L_{3})/66420043,\\
   \indent f6[11]=(4761505236L_{2}+406049487 L_{3}
           -531360344L_{5})/265680172,\\
   \indent f6[12]=3(-730884597L_{2}+350022565L_{3})/1461240946,\\
   \indent f6[13]=3(-1023320876L_{2}+227882885L_{3})/66420043,\\
   \indent f6[14]=3 (72905938844L_{2}-1923836317 L_{3}
           -5313603440 L_{5})/664200430,\\
    \indent f6[16]=L_{1},\indent f6[17]=123(-20616556L_{2}+22289L_{3})/265680172,\\
      \indent f6[18]= 3 (8124502060L_{2}-245230307 L_{3}
           -531360344L_{5})/265680172,\\
   \indent    f6[22]=L_{1},\indent f6[23]=-\frac{1}{2} L_{1},\indent f6[24]=L_{1},\indent f6[25]=L_{2},\\
   \indent f6[26]=(-11324245500 L_{2}-806175417L_{3}
           +1062720688 L_{5})/132840086,\indent f6[27]=-\frac{1}{6} L_{1},\\
   \indent f6[28]=(-13043019308 L_{2}+255536149 L_{3}+
           996300645 L_{4}-664200430 L_{6})
         /664200430,\\
    \indent f6[29]=L_{3},\indent f6[30]=(26247460620 L_{2}-728072793 L_{3}-1062720688 L_{5})/
           265680172,\\
     \indent f6[31]=\frac{2}{3} L_{1},\indent f6[32]=L_{4},\indent f6[34]=L_{1}, \indent f6[35]=\frac{2}{3} L_{1},\\
 \indent f6[37]=(-23291863284 L_{2}-2029012023 L_{3}+2656801720 L_{5})/
           1328400860, \\
    \indent f6[38]=L_{5},  \indent f6[39]=L_{6},\indent f6[40]=\frac{1}{3} L_{1},\\
   \indent f6[42]=(21371821812 L_{2}-2351490636 L_{3}- 1660501075
           L_{4}+1328400860 L_{6})/6642004300,\\
   \indent f6[43]= 2(-858895785 L_{2}-3491642 L_{3}+66420043
           L_{6})/332100215,$\\
\noindent where $L_{1},L_{2},L_{3},L_{4},L_{5},L_{6}$ are arbitrary constants but do not vanish at the same time.\\

\noindent {\bf V Exact Spinor Field Constructions of $W_{6}$ String}\\
\indent Using our program we also checked the results of the
$W_{2,3}$ string and
    the $W_{2,5}$ string, and got a more general solution of the $W_{2,4}$ string.
    Furthermore, we find the solutions of $W_{2,s}(s=3,4,5,6)$ strings are
    very standard, that is, there are three solutions for each '$s$'. In these solutions,
    the values of '$y$' correspond to '0','1' and an arbitrary constant respectively.
    Thus $Q_{N}(N=s-2)$ can be either of $Q_{N}^{(1)}$, $Q_{N}^{(2)}$ and $Q_{N}^{(3)}$.
    We standardize these solutions as follows: $Q_{N}^{(1)}$ corresponds to the solution of
    $y=0$, $Q_{N}^{(2)}$ to $y=1$, and $Q_{N}^{(3)}$ corresponds
    to the case that $'y'$ is an arbitrary constant. By this assumption, we rewrite the results
    of $W_{2,3}$ string and give out the new constructions of $W_{2,4}$ string.
    The solutions of $W_{2,5}$ string are the same with that in Ref.[7].\\

    \noindent{\bf V.1 The solutions of $W_{2,3}$ string}\\
    \indent      $ F3=f3[1] \beta^3\gamma^3+f3[2]\beta\gamma^2
                    \partial\beta + f3[3]\partial \beta \partial \gamma
                    +f3[4] \beta \gamma  \psi\partial \psi + f3[5] \beta
                    \partial ^2\gamma +f3[6] \psi  \partial ^2\psi
                 $.\\[2mm]
    \noindent   (1)  $ y=0 $ and \\
    \indent         $ f3[4]=f3[6]=0 $,\\
    \noindent   where other coefficients are arbitrary constants but do not vanish at the same time.\\[2mm]
    \noindent   (2) $ y=1 $ and\\
    \indent         $f3[1]  = \frac{1}{150} (-7 N_{1}+3N_{2}),\indent  f3[2]=\frac{1}{15}(7 N_{1}-3 N_{2}) , \indent f3[3]=N_{1},$\\
    \indent         $f3[4] = -\frac{2}{5}(11 N_{1}-39 N_{2}),\indent  f3[5]=N_{2}, \indent f3[6]=11N_{1}-39N_{2} ,$ \\
    \noindent   where $N_{1}$ and $N_{2}$ are arbitrary constants but do not vanish at the same time.\\[2mm]
    \noindent   (3) $y$ is an arbitrary constant  and\\
    \indent         $ f3[4]=f3[6]=0, \indent f3[1]=-8 P, \indent f3[2]=80 P,\indent f3[3]=195 P,\indent f3[5]=P  $,\\
    \noindent   where $P$ is an arbitrary nonzero constant. \\

    \noindent{\bf V.2 The new solutions of $W_{2,4}$ string}\\
   \indent    $F4  = f4[1]\beta ^4\gamma ^4+f4[2](\partial \beta)^2\gamma ^2+f4[3] \beta ^3\gamma ^2\partial \gamma + \ f4[4]\beta^2 (\partial \gamma )^2+f4[5] \beta ^2\gamma ^2\psi \partial \psi \\
    \indent\indent       +f4[6]\partial \beta \gamma  \psi  \partial \psi + f4[7]\beta \partial \gamma  \psi  \partial \psi +f4[8] \beta  \partial ^2\beta  \gamma ^2+f4[9] \partial ^2\beta \partial \gamma + \ f4[10]\partial \beta \partial ^2\gamma  \\
     \indent\indent            +f4[11] \partial \psi \partial ^2\psi +f4[12] \beta \partial ^3\gamma + f4[13]\psi  \partial ^3\psi .\\
               $
    \noindent   (1)  $ y=0 $ and \\
  \indent $ f4[5]=f4[6]=f4[7]=f4[11]=f4[13]=0 ,$\\
    \noindent   where other coefficients are arbitrary constants but do not vanish at the same time.\\[2mm]
    \noindent   (2) $ y=1 $ and\\
   \indent    $f4[1] =\frac{1}{468930}(-58 Z_{1}-114 Z_{2}+18207 Z_{3}),\indent f4[2] =\frac{1}{15312}(-116 Z_{1}-63 Z_{2}+15426 Z_{3}), \\
   \indent              f4[3]=\frac{70}{3} f4[1],\indent f4[4] =\frac{1}{232}(-5 Z_{2}+114 Z_{3}), \indent f4[5] =\frac{1}{21} (-4 Z_{1}+3 Z_{2}), \\
   \indent            f4[6]=Z_{1},\indent f4[7]=Z_{2},\indent f4[8]=Z_{3},\indent f4[9]  =\frac{1}{696} (-7 Z_{2}+1134 Z_{3}+232 Z_{4}),  \\
   \indent         f4[10]=Z_{4},\indent f4[11] =\frac{1}{7} (33 Z_{1}+Z_{2}),\indent f4[12]=\frac{35}{1392} (Z_{2}-798 Z_{3}+928Z_{4}),\\
   \indent         f4[13]=\frac{1}{21} (-11 Z_{1}+16 Z_{2}), $\\
    \noindent   where $Z_{1},Z_{2},Z_{3}$ and $Z_{4}$ are arbitrary constants
                    but do not vanish at the same time.\\[2mm]
    \noindent   (3) $y$ is an arbitrary constant  and\\
      \indent $    f4[5]=f4[6]=f4[7]=f4[11]=f4[13]=0 ,\indent f4[1]=\frac{867}{22330} V_{1}, \indent f4[2]=\frac{2571}{2552} V_{1},\\
       \indent f4[3]=\frac{289}{319} V_{1}, \indent f4[4]=\frac{57}{116} V_{1} ,\indent f4[8]=V_{1},\indent f4[9]=\frac{189}{116} V_{1}+\frac{1}{3} V_{2},\\
       \indent f4[10]=V_{2},\indent f4[12]=-\frac{133}{232} V_{1} +\frac{2}{3} V_{2} ,$\\
    \noindent   where $V_{1}$ and $V_{2}$ are arbitrary constants but do not vanish at the same
                time.\\
    \indent Comparing these new results with that in Ref.[10], we found they are more general.\\

\noindent{\bf V.3 Exact spinor field constructions of $W_{6}$ string}\\
\indent In this case the $BRST$ charge can be written as
    $Q_B=Q_0+ Q_1+ Q_2+Q_3+ Q_4$, where $Q_i(i=1,2,3,4)$ must be
    one of the spinor field realization
    of $W_{2,s}(s=3,4,5,6)$ string respectively.
    The selection rule equation(10) namely
    $\{Q_1, Q_2\}=\{Q_1, Q_3\}=\{Q_1, Q_4\}=\{Q_2, Q_3\}=\{Q_2,
    Q_4\}=\{Q_3, Q_4\}=0$. For these selection rules, we list two
    tables in which the correct combinations are denoted as sign '$\surd$' whilst
    the incorrect combinations are denoted as sign '$\times$'. Then the exact
    constructions of the spinor field grading $BRST$
    charge for the $W_{6}$ string are obtained. Especially we give out two simple correct
    combinations here.\\
    \\
    \noindent{\bf V.3.1 The simplest combination of}
    $Q_{0},Q_{1}^{(1)}, Q_{2}^{(1)}, Q_{3}^{(1)}$ and
    $Q_{4}^{(1)}$
        $$\align
            Q_{B}=\oint dz \; & [ \; c(-\frac{1}{2}\eta_{\mu\nu}\partial Y^{\mu}Y^{\nu}
                - \frac{1}{2}\partial\psi\psi+2K b\partial c+K\partial bc)\\
                & + f_{1}\beta_{1}^3\gamma_{1}^3 + f_{2}\beta_{2}^4\gamma_{2}^4
                + f_{3}\beta_{3}^5\gamma_{3}^5 + f_{4}\beta_{4}^6\gamma_{4}^6 \; ]
        \endalign $$
    \indent\indent   where $f_{i}(i=1,2,3,4)$ are arbitrary nonzero constants. \\
    \indent\indent This is one of the simplest formulae of $Q_{B}$.\\

    \noindent {\bf V.3.2 Another simple combination of}
        $Q_{0},Q_{1}^{(3)}, Q_{2}^{(3)}, Q_{3}^{(3)}$ and
        $Q_{4}^{(3)}$
        $$\align
            Q_{B} = \oint dz\;
                    & [\; c(-\frac{1}{2}\eta_{\mu\nu}\partial Y^{\mu}Y^{\nu}
                    - \frac{1}{2}\partial \psi \psi + 2K b\partial c + K \partial bc
                    + 3\beta_{1}\partial\gamma_{1}+2\partial\beta_{1}\gamma_{1}\\\allowdisplaybreak
                    &  + 4\beta_{2}\partial\gamma_{2}+3\partial\beta_{2}\gamma_{2}
                    + 5\beta_{3}\partial\gamma_{3}+4\partial\beta_{3}\gamma_{3}
                    + 6\beta_{4}\partial\gamma_{4}+5\partial\beta_{4}\gamma_{4})\\\allowdisplaybreak
                    & + g_{1}(-8 \beta_{1}^3\gamma_{1}^3
                    + 80\beta_{1}\gamma_{1}^2 \partial\beta_{1}
                    + 195\partial \beta_{1} \partial \gamma_{1}
                    + g_{1} \beta_{1} \partial^2\gamma_{1})\\\allowdisplaybreak
                    & + g_{2}( \partial ^2\beta_{2} \partial \gamma_{2}
                    + 3 \partial\beta_{2} \partial ^2\gamma_{2}
                    + 2 \beta \partial_{2} ^3\gamma_{2})\\\allowdisplaybreak
                    & + g_{3}(3\partial^{2}\beta_{3}\partial^{2}\gamma_{3}
                    + \partial\gamma_{3}\partial^{(3)}\beta_{3}
                    + 2\partial\beta_{3}\partial^{(3)}\gamma_{3})\\\allowdisplaybreak
                    & + g_{4}(6 \partial^3\beta_{4}\partial^2\gamma_{4}
                    + 4 \partial^2\beta_{4}\partial^3\gamma_{4}
                    - \partial^5\beta_{4}\gamma_{4})
                    \;]
        \endalign $$
\indent\indent   where $g_{i}(i=1,2,3,4)$ are any arbitrary nonzero constants. \\

\noindent {\bf V.3.3 The complete combinations of
$Q_{i}(i=1,2,3,4)$}

\indent The exact combinations could be found in Table 1 and Table
2.

{\begin{center}
\begin{center}{ Table 1:
The precise combinations of $Q_{1}$,$Q_{2}$,$Q_{3}$ and $Q_{4}$}
\end{center}
    \renewcommand\arraystretch{1.5}
    \begin{tabular}{|c|c|c|c|c}
        \hline \hline
        combinations  & $Q_{2}^{(1)}\;Q_{2}^{(2)}\;Q_{2}^{(3)}$
                      & $Q_{3}^{(1)}\;Q_{3}^{(2)}\;Q_{3}^{(3)}$
                      & $Q_{4}^{(1)}\;Q_{4}^{(2)}\;Q_{4}^{(3)}$\\
        \hline
        $Q_{1}^{(1)}$  &  $\surd \;\;\;\;\; \surd \;\;\;\;\; \surd$         &  $\surd \;\;\;\;\; \surd \;\;\;\;\; \surd$       &  $\surd \;\;\;\;\; \surd \;\;\;\;\; \surd$             \\
        $Q_{1}^{(2)}$  &  $\surd \;\;\;\;   \times^1 \;\;\;\;\surd$         &  $\surd \;\;\;\;   \times^2\;\;\;\; \surd$       &  $\times^*  \;\;   \times^4\ \;\;   \times^*$          \\
        $Q_{1}^{(3)}$  &  $\surd \;\;\;\;\; \surd \;\;\;\;\; \surd$         &  $\surd \;\;\;\;\; \surd \;\;\;\;\; \surd$       &  $\surd \;\;\;\;\; \surd \;\;\;\;\; \surd$             \\
        \hline
        $Q_{2}^{(1)}$  &  $\; - \;\;\;\; - \;\;\;\; - \;$                   &  $\surd \;\;\;\;\; \surd \;\;\;\;\; \surd$       &  $\surd \;\;\;\;\; \surd \;\;\;\;\; \surd$             \\
        $Q_{2}^{(2)}$  &  $\; - \;\;\;\; - \;\;\;\; - \;$                   &  $\surd \;\;\;\;   \times^3\;\;\;\; \surd$       &  $\times^*  \;\;   \times^5\ \;\;   \times^*$          \\
        $Q_{2}^{(3)}$  &  $\; - \;\;\;\; - \;\;\;\; - \;$                   &  $\surd \;\;\;\;\; \surd \;\;\;\;\; \surd$       &  $\surd \;\;\;\;\; \surd \;\;\;\;\; \surd$             \\
        \hline
        $Q_{3}^{(1)}$  &  $\; - \;\;\;\; - \;\;\;\; - \;$                   &  $\; - \;\;\;\; - \;\;\;\; - \;$                 &  $\surd \;\;\;\;\; \surd \;\;\;\;\; \surd$             \\
        $Q_{3}^{(2)}$  &  $\; - \;\;\;\; - \;\;\;\; - \;$                   &  $\; - \;\;\;\; - \;\;\;\; - \;$                 &  $\times^*  \;\;   \times^6\ \;\;   \times^*$          \\
        $Q_{3}^{(3)}$  &  $\; - \;\;\;\; - \;\;\;\; - \;$                   &  $\; - \;\;\;\; - \;\;\;\; - \;$                 &  $\surd \;\;\;\;\; \surd \;\;\;\;\; \surd$             \\
        \hline
    \end{tabular}
    \vskip 10pt
\end{center}

$\;$

\begin{center}
\begin{center}{ Table 2:
The special combinations of $Q_{i}$ and $Q_{j}$ when $f6[31]=0$}
\end{center}
    \renewcommand\arraystretch{1.5}
    \begin{tabular}{|c|c|c}
        \hline \hline
        combinations   &   $\;Q_{j}^{(1)}\;\;Q_{j}^{(2)}\;\;Q_{j}^{(3)}\;$   \\
        \hline
        $Q_{i}^{(1)}$  &  $\surd \;\;\;\;\; \surd  \;\;\;\;\; \surd$     \\
        $Q_{i}^{(2)}$  &  $\surd \;\;\;\;\; \times \;\;\;\;\; \surd$     \\
        $Q_{i}^{(3)}$  &  $\surd \;\;\;\;\; \surd  \;\;\;\;\; \surd$     \\
        \hline
    \end{tabular}\\
\end{center}

 \noindent $^1$If we take $f4[6]=f4[7]=0$(we name it condition 1) , this combination will be correct. \\
             \noindent $^2$If we take $f5[8]=f5[9]=f5[14]=0$(we name it condition 2), this combination will be correct.\\
             \noindent $^3$If we take condition 1 or 2, this combination will be correct.\\
             \noindent $^4$If we take $f6[6]=f6[9]=f6[10]=f6[15]=f6[20]=f6[25]=f6[29]=f6[31]=0$(we name it condition 3), this combination will be correct.\\
             \noindent $^5$If we take condition 1 or 3, this combination will be correct.\\
             \noindent $^6$If we take condition 2 or 3, this combination will be correct.\\
             \noindent $^*$If we take $f6[31]=0$ , all the '$\times^*$'  in Table 1 will become '$\surd$'.
             Then each part of Table 1 changes to Table 2.\\
}

\noindent{\bf VI Conclusion}\\
\indent In conclusion, the spinor field grading $BRST$ charges
         of $W_{2,6}$ string and $W_{6}$ string
    have been constructed. And a special program was obtained to construct spinor field
    $BRST$ charges of a general $W_{2,s}$ strings  by using the $BRST$ method. With this procedure, the results are more
    accurate, and the process is accelerated rapidly. We have checked the
    solutions of $W_{2,3}$ string as well as $W_{2,5}$ string and given out a more general
    result of $W_{2,4}$ string in spinor field. Observing these
    solutions, we found they are standard. Of course the more realizations of
    $W_{2,s}$ strings can be calculated by using our program. Using these results, we will discuss their physical states in our next
    work.\\

\noindent{\bf Acknowledgements}\\
\indent We are grateful to Professor C.N.Pope as well as
        Dr. H.Wei for their useful discussion. We have used the Mathematica's
         package OPEdefs.m written by Professor Kris
         Thielemans[12].\\

\end{document}